# Spatial population expansion promotes the evolution of cooperation in an experimental Prisoner's Dilemma

J. David Van Dyken*[12], Melanie J.I. Müller[23], Keenan M.L. Mack[4], Michael M. Desai[123]
[1]Department of Organismic and Evolutionary Biology, Harvard University,
[2]FAS Center for Systems Biology, Harvard University
[3]Department of Physics, Harvard University
[4]National Institute for Mathematical and Biological Synthesis, University of Tennessee, USA
*vandyken@fas.harvard.edu

## Summary

**Cooperation is ubiquitous in nature, but explaining its existence remains a central interdisciplinary challenge [1-3]. Cooperation is most difficult to explain in the Prisoner's Dilemma (PD) game, where cooperators always lose in direct competition with defectors despite increasing mean fitness [1, 4, 5]. Here we demonstrate how spatial population expansion, a widespread natural phenomenon [6-11], promotes the evolution of cooperation. We engineer an experimental PD game in the budding yeast *Saccharomyces cerevisiae* to show that, despite losing to defectors in non-expanding conditions, cooperators increase in frequency in spatially expanding populations. Fluorescently labeled colonies show genetic demixing [8] of cooperators and defectors followed by increase in cooperator frequency as cooperator sectors overtake neighboring defector sectors. Together with lattice-based spatial simulations, our results suggest that spatial population expansion drives the evolution of cooperation by 1) increasing positive genetic assortment at population frontiers and 2) selecting for phenotypes maximizing local deme productivity. Spatial expansion thus creates a selective force whereby cooperator-enriched demes overtake neighboring defector-enriched demes in a "survival of the fastest". We conclude that colony growth alone can promote cooperation and prevent defection in microbes. Our results extend to other species with spatially restricted dispersal undergoing range expansion, including pathogens, invasive species and humans.**

## Highlights

**-Spatial population expansion selects for cooperation at population frontiers**
**-Frontiers possess high genetic relatedness and create among-group competition favoring "survival of the fastest"**
**-Spatial expansion favors genotypes promoting maximization of group productivity**
**-Colony growth in microbes is a force promoting cooperation**

Cooperation forms the basis for numerous complex phenotypes, from cell-cell communication and biofilm formation in microbes to nest construction in multicellular species [12, 13]. Explaining how cooperation evolves despite the direct fitness advantage gained by free-riding remains a central challenge in biology and the social sciences [1-3]. This is particularly true in the Prisoner's Dilemma game [1, 4, 5] where cooperators always lose in direct competition with defectors, leading to a "Tragedy of the Commons" [14]. Nearly half a century of research on social evolution has offered insight into this dilemma, with inclusive fitness theory focusing attention on the importance of high genetic relatedness [15], multilevel selection theory highlighting the role played by competition among social groups [16, 17], and spatial games showing the importance of population structure [18-21]. Here we bring these three frameworks together, demonstrating that spatial population expansion creates a setting where both relatedness and intergroup competition are amplified, promoting the evolution of cooperation.

Spatial population expansion occurs when a species spreads outward to fill vacant eco-space or to supplant resident species, resulting in increased geographic range. Human migration out of Africa [7] is one salient example, while more generic examples include ecological invasions, epidemics, growth of microbial colonies, and expansion due to habitat modification caused by environmental disturbance or climate change [8-11]. Spatially expanding populations form a propagating density wave called a "Fisher wave" with a constant speed proportional to the square root of the mean growth rate (i.e., Malthusian fitness) of subpopulations at the front [22]. Genetically heterogeneous populations may also form an "allele frequency" wave representing the spatial spread of alleles [22] (see Supplemental Information).

A small number of colonists initiate new subpopulations at the leading edge of the propagating Fisher wave, creating a repeated series of genetic bottlenecks or "founder effects" that cause stochastic loss of genetic diversity at frontiers [23, 24]. Because cooperation is favored under conditions of high genetic relatedness [15], range expansions could thus in principle favor the evolution of cooperation. However, there are at least three complicating factors. First, within-subpopulation selection favoring defection opposes genetic demixing, potentially preventing high cooperator relatedness from ever arising. Second, even with high relatedness it is not clear what selective force, if any, favors cooperation over defection in expanding populations. Finally, almost any genotype that stochastically fixes at the front can increase in frequency via "surfing" [6, 10, 24, 25]. Other genotypes that stochastically fall behind this front cannot typically catch up, even if they are more fit. This is because they expand outwards in a trailing allele frequency wave traveling at a speed determined by the difference in fitness between defector and cooperator genotypes, ($W_D - W_C$), which will often be much smaller than mean absolute fitness. For social traits, a genotype fixed at the frontier will outrun genotypes in the population interior and increase in global frequency provided that $(1+b)/2 > c$, where $b$ is the social and $c$ the direct fitness effect of the leading genotype (Supplemental Information). The social effect here refers to the fitness increment or decrement received by an individual from social partners (e.g. the benefit of the public good), while the direct effect is the fitness increment or decrement accrued to an individual for engaging in a social behavior (e.g. the cost of producing the public good). Note that this condition $(1+b)/2 > c$ can be satisfied even when $b < 0$, and thus in principle surfing may promote cooperation's opposites, selfishness and spite [26], including spite against relatives. Given these complications, it is not clear whether spatial expansion will in fact promote the evolution of cooperation.

To test the effect of spatial expansion on defector/cooperator dynamics, we engineered an experimental Prisoner's Dilemma game using cooperative sucrose metabolism in haploid, vegetatively growing strains of the budding yeast, *Saccharomyces cerevisiae* [27]. Yeast secrete the exo-enzyme invertase in order to digest the disaccharide sucrose, which cannot easily be imported into the cell, forming monosacharides that are readily imported. In our strains, sucrose cannot be imported at all due to disruption of the genes *mal12* and *mal22* [28]. Because digestion occurs externally, invertase producers ("cooperators") create a public good that is exploitable by non-producers ("defectors"), who gain a relative fitness advantage by not paying the fitness cost of production [27, 29]. We engineered a fluorescently marked defector strain by deleting the invertase gene *SUC2*.

We note that in minimal sucrose media (YNB + 2% sucrose), competitions between *SUC2*$^{+}$ and *suc2*$^{-}$ strains in shaken liquid culture were previously found to follow Snowdrift game dynamics [29]. In a Snowdrift game, the rare type (regardless of whether it is a cooperator or a cheater) has a fitness advantage, leading to stable maintenance of both cooperators and defectors [1, 30]. The maintenance of cooperation is therefore easily ensured, in contrast to the Prisoner's Dilemma game where the maintenance of cooperation is much more difficult to explain [1, 30]. In addition, cooperators in a Snowdrift game have a colonization advantage over defectors since defectors cannot colonize habitat unoccupied by cooperators [28]. This conflates colonization ability and cooperation by linking both to a single genotype. Because spatial expansion is already known to select for colonization ability [31, 32], such linkage would prevent us from concluding that spatial expansion favors cooperation per se rather than superior colonization ability. By contrast, defectors in a Prisoner's Dilemma game do not require the presence of cooperators to colonize new habitat, making it possible to disentangle selection for cooperation from colonization ability.

We therefore used two approaches to construct a Prisoner's Dilemma from this system. First, we eliminated the rare advantage of cooperators that is necessary for Snowdrift dynamics by conducting competitions in media (YEP + 2% sucrose) in which our defector strains could grow in the absence of cooperators (most likely by consuming amino acids available in YEP, although growth is slower than for cooperators; Fig. 1A). This environment also eliminates the difference in colonization ability between cooperator and cheater strains, as cheaters no longer require the presence of cooperators to colonize the frontier (green line in Fig. 1A). Next, we engineered a defector strain that is resistant to cycloheximide, a translation-inhibiting drug that limits growth by binding to ribosomal subunit *cyh2*. This creates a system in which we can experimentally impose a tunable "cost of cooperation" by varying the level of cycloheximide in the growth media. Specifically, increasing the cycloheximide concentration slows the growth of cooperators but not the resistant defectors, leading to an increased "cost of cooperation."

When mixed with our defector strain in an unstructured environment (shaken liquid culture), our cooperator strain declines at all frequencies when a cost of cooperation is imposed, despite having a superior growth rate over defectors in pure culture (Figs. 1, S3). These results are consistent with Prisoner's Dilemma evolutionary dynamics. Unlike in a Snowdrift game, any increase in frequency of cooperators in our experiments is not due to rare cooperator advantage.

To determine whether spatial expansion can promote cooperation in our experimental Prisoner's Dilemma, we initiated spatial expansions by spotting a droplet of mixed cooperator/defector cultures onto solid media (YEP + 2% sucrose + 2% agar) for a range of imposed costs (see Experimental Procedures). Spatial diffusion of cells in *S. cerevisiae* is caused when cellular growth generates an outward force leading to radial spatial expansions of colonies [8]. Note that expansion is not caused by active cell motility in this system as yeast lack motility. Relative frequency measurements taken using flow cytometry show that cooperators initially declined in frequency at a rate consistent with that of well-mixed liquid competitions, but then increased in frequency as expansion proceeds (Fig. 2C). Likewise, image analysis of fluorescently labeled colonies shows low cooperator frequency near the initial site of inoculation (the "homeland"), but then increasing frequency with increasing distance from the homeland (Fig. 2E). Lattice-based spatial simulations of a Prisoner's Dilemma show the same spatio-temporal dynamic of initial decline in cooperator frequency followed by increase as expansion proceeds (Fig. 2D). Cooperators invade when rare over a range of imposed costs (Fig. 3). Furthermore, when the benefit of cooperation is removed by competing strains on glucose media, the cooperator strain no longer increases in frequency upon spatial expansion (Fig. S4). These data clearly demonstrate that spatial expansion can promote the evolution of cooperation.

How does spatial expansion promote cooperation? Fluorescent colony images reveal the formation of discrete sectors of fixed genotypes (Figs. 2A,B, 3), which is the colony-level signature of genetic demixing [8]. Thus, spatial expansion can lead to high positive assortment of cooperators via genetic demixing, even though this assortment is opposed by selection favoring defectors within demes (Fig. 1B, Supplemental Information). Local fixation of cooperators at frontiers despite counterselection within demes is analogous to surfing by deleterious mutations [6, 10, 24, 25] and requires similar conditions to obtain. In other words, spatial expansion leads to the formation of uniform sectors of cooperators or defectors, increasing genetic relatedness of nearby individuals. This diminished local genetic diversity reduces the direct competition between cooperators and defectors (see also [33]), thereby mitigating the principle selective advantage of defection.

We note that genetic demixing (i.e. “sectoring”) is particularly clear in our experimental yeast system because yeast lack motility and “dispersal” of offspring is local. In other systems, movement of individuals and dispersal of offspring can in principle blur sector boundaries and oppose demixing at the frontier. In the extreme case where movement and dispersal are very long-range, the spatial sectoring we describe here will not occur, and our analysis or results will not generalize to this situation. However, in real populations movement and dispersal are usually spatially restricted: a migrant is more likely to disperse nearby than far away. In this case, genetic demixing will occur provided that outward range expansion is sufficiently rapid compared to the rate of dispersal between occupied demes across sector boundaries (i.e., perpendicular to the expansion direction) [24, 25, 34]. In nature, species as diverse as rabies virus [35] and humans [36] show genetic signatures of expansion-associated demixing and sectoring, suggesting that the phenomenon we describe here may apply more generally. To the extent that spatial expansion-associated genetic demixing is possible in a species, the mechanism we describe here promoting the evolution of cooperation will also be possible.

Once cooperator sectors establish, their overall productivity will be higher than that of defector sectors provided the fitness benefit of cooperation exceeds the cost of cooperation, $b > c$ (Supplementary Information). When this is true, cooperator sectors will expand radially faster than neighboring defector sectors, leading to a corresponding expansion of the boundaries of the cooperator sectors at the expense of neighboring defector sectors as we see in our experiements (Figs 2BE, 3). This leads to an overall increase in cooperator frequency, and suggests that range expansion creates a force of natural selection favoring the "survival of the fastest." This force acts to promote genotypes supporting maximal group productivity, since high productivity sectors expand at a faster rate, allowing them to overtake lower productivity sectors.

We turn to stochastic, lattice-based spatial simulations to further test the survival of the fastest hypothesis. To test whether survival of the fastest is indeed necessary for range expansion to promote cooperation, we eliminate this force by restricting expansion to one dimension in our spatial simulations. In one dimension, an expanding subpopulation has no neighboring subpopulations to compete with so that intergroup competition is absent. In this case, we find that the probability of cooperator establishment at the front is never greater than the neutral probability of establishment (which is equal to the initial frequency of the allele, $p_0$) and declines with increasing cost (Fig. 4). Put differently, cooperators can only outrun defectors in a one-dimensional

Prisoner's Dilemma if they randomly take over the frontier, an outcome uniformly opposed by selection (Supplemental Information).

In contrast to the one-dimensional case, the probability of cooperators fixing at the frontier is substantially higher in two dimensions, where subpopulations compete with neighbors for occupancy of uncolonized habitat (Fig. 4). Two-dimensional expansions are also more efficient at purging deleterious alleles from frontiers, as seen in comparison of the black and gray dashed lines in Fig. 4. Our data support the conclusion that two-dimensional spatial expansions generates selection at the frontier for genotypes that maximize group productivity, as these genotypes lead to the greatest expansion velocity of the front, allowing cooperator enriched demes to overtake defector enriched demes. Spatial expansion generates both conditions necessary for natural selection: heritability (positive assortment of social strategies, making one's social environment heritable) and differential success (survival of the "fastest").

Microbes posses a multiplicity of cooperative phenotypes [12], and rapid cell division in conjunction with large colony sizes makes the repeated emergence of defector mutants inevitable in nature. We have demonstrated that colony growth itself creates a force that promotes cooperation and inhibits colony invasion by defector mutants. Range expansions may promote cooperation more generally and may allow already cooperative species to shed social parasites, so long as the pattern and rate of dispersal and reproduction allow for genetic demixing upon expansion. It is possible that reduced cheater load upon expansion may accelerate biological invasion by cooperative species, with potential implications for biological control**.** Stochastic demixing may also occur with culturally transmitted phenotypes, such that range expansion may have been important in the spread of cultural norms facilitating cooperation in humans. Yet this force persists only as long as expansion continues. Repeated cycles of expansion and contraction, possibly due to frequent disturbance, may be necessary to maintain persistent selection for cooperation by this mechanism.

**Acknowledgements** We thank John Koschwanez for strains, and Michael Whitlock, Benjamin Good, Michael McDonald, Lauren Nicolaisen, Katya Kosheleva, John Koschwanez and members of the Desai lab for many useful discussions. JDVD acknowledge support from a National Science Foundation Postdoctoral Fellowship. KMLM acknowledges support from the National Institute for Mathematical and Biological Synthesis, an Institute sponsored by the National Science Foundation, the U.S. Department of Homeland Security, and U.S. Department of Agriculture through NSF Award #EF-0832858, with additional support from the University of Tennessee, Knoxville. MMD acknowledges support from the James S. McDonnell Foundation and the Alfred P. Sloan Foundation.

## FIGURE LEGENDS

**Figure 1| An experimental Prisoner's Dilemma. A)** Populations composed of all cooperators (red) have a higher growth rate than pure defector populations (green), but **B)** cooperators lose to defectors within mixed populations. Growth rate in A) was assayed on agar plates by measuring colony radius over time, which is directly proportional to rate of cell division *S. cerevisiae* [8]. Lines in B) represent cooperator frequency trajectories, measured with FACS, in shaken liquid culture over the course of one week for four different levels of imposed cost (cycloheximide concentrations: 50, 75, 100, and 150nM, from top to bottom and blue to red).

**Figure 2| Spatial expansion promotes the evolution of cooperation in a Prisoner's Dilemma. A)** Growth of fluorescently labeled colonies (cooperators in red, defectors green). **B)** Competitions inoculated with different initial cooperator frequencies (from top to bottom: 0.99, 0.90, 0.50, 0.10, and 0.01) after 7 days of growth (not to scale). Note visible expansion of cooperator (red) sectors at colony frontiers, and proliferation of defectors (green) in colony interior. **C)** Frequency trajectory of the cooperator strain in spatially expanding (solid lines) and stationary (non-expanding) (dotted lines) competitions as measured by flow cytometry (FACS) over the experiment. Dotted lines follow the same populations over the whole time-course, while spatial expansions required destructive sampling of colonies at each time point. Note that frequency is measured over whole colony, not just at frontier. **D)** Frequency dynamics from lattice-based spatial simulations of a PD game with non-overlapping generations in radially expanding (solid lines) and stationary (dotted lines) populations initiated from well-

mixed (relatedness = 0) homelands of varying initial cooperator frequencies, with simulation parameters: $W_0 = 1$, $K = 50$, $m = 0.2$, $b = 0.5$, $c = 0.1$ (see Methods). **E)** Image analysis of experimental colonies after 7 days of growth. Cooperator frequency measured along the circumference of a circle of radius r centered at the colony center (importantly, this means that frequency is not cumulative). Each line denotes the average over 3 replicates for different imposed cost (cycloheximide concentrations 0nM [dashed blue line] to 200nM [solid red line] at 25nM increments) with initial cooperator frequency of 0.10. Imposed cost in A,B,C was 50nM cycloheximide.

**Figure 3| Invasion of rare cooperators during colony expansion over a range of imposed costs.** Colonies imaged after 7 days of growth (not to scale). Competitions inoculated with cooperators at frequency 0.10 (top row) or 0.01 (bottom row), at different cycloheximide concentrations.

**Figure 4| Selection for cooperation in spatially expanding populations requires competition among neighboring frontier subpopulations.** Results of lattice-based spatial simulations of a PD game in a population expanding in one direction (i.e., a linear front[8]). **A)** Example of the endpoint of a two-dimensional simulation (cooperators in red, defectors in green), while **B)** shows averages over 100-500 iterations, for two-dimensional (black) and one-dimensional (grey) simulations. Each lattice site is a subpopulation growing logistically to size $K = 50$ in a metapopulation of dimension 1x150 (one-dimensional) or 25x150 (two-dimensional) sites. The vertical axis in **B)** gives the frequency of cooperators at the population frontier (defined here as a 1x10 or

25x10 area at the furthest edge of the population) after 200 generations. Horizontal dotted line indicates neutral expectation of this value. Simulations initiated with cooperators at 0.10 frequency in a well-mixed (relatedness = 0) homeland of length 10 sites. Two-dimensional expansions (black) select for cooperation, but one-dimensional expansions (grey) cannot. $b$ = 0.5 for solid lines, and dashed lines denote zero social effect ($b$ = 0); note that the horizontal axis has been normalized by $b$ = 0.5. Simulation parameters: $W_0$ = 1, $K$ = 50, $m$ = 0.2.

Figure 1

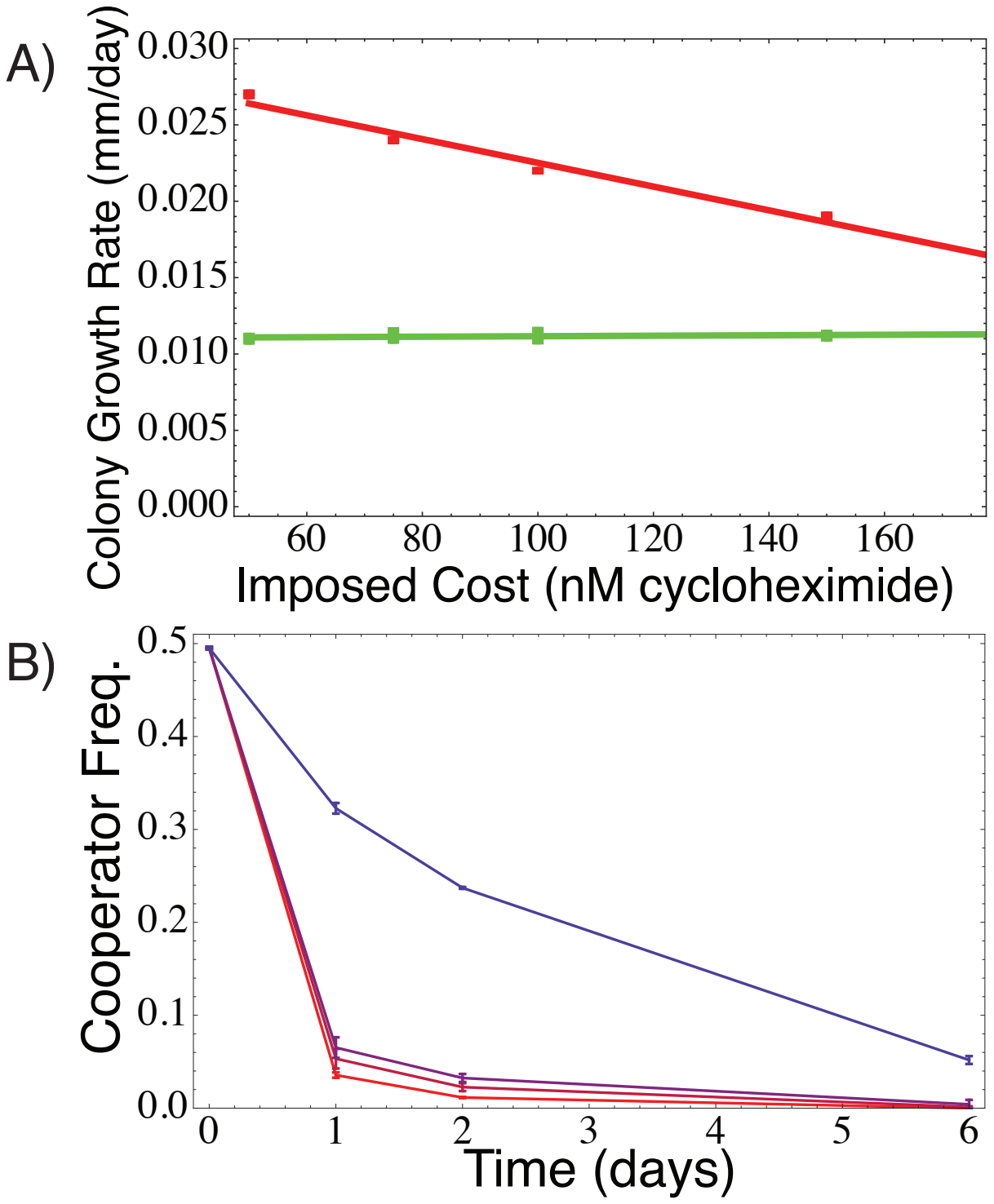

Figure 2

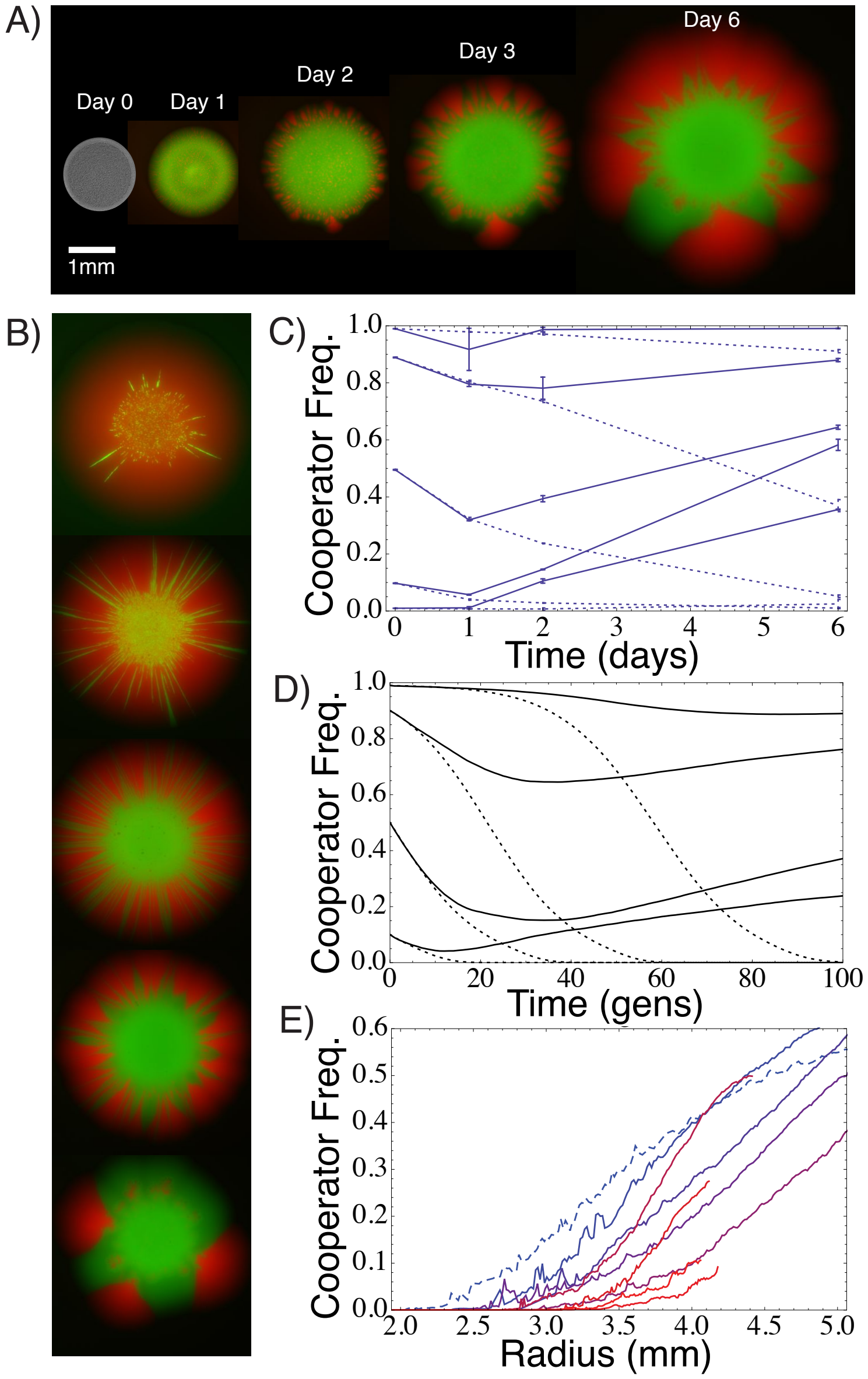

Figure 3

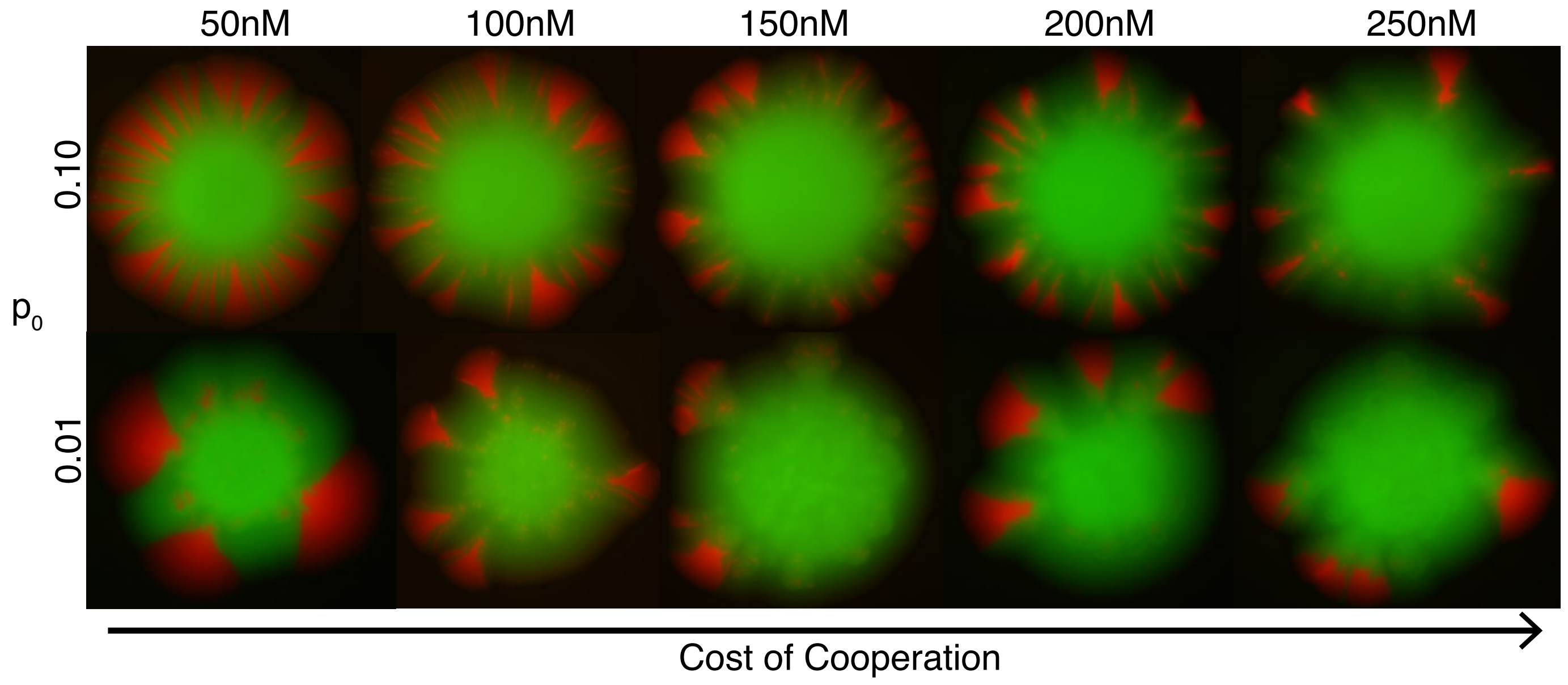

Figure 4

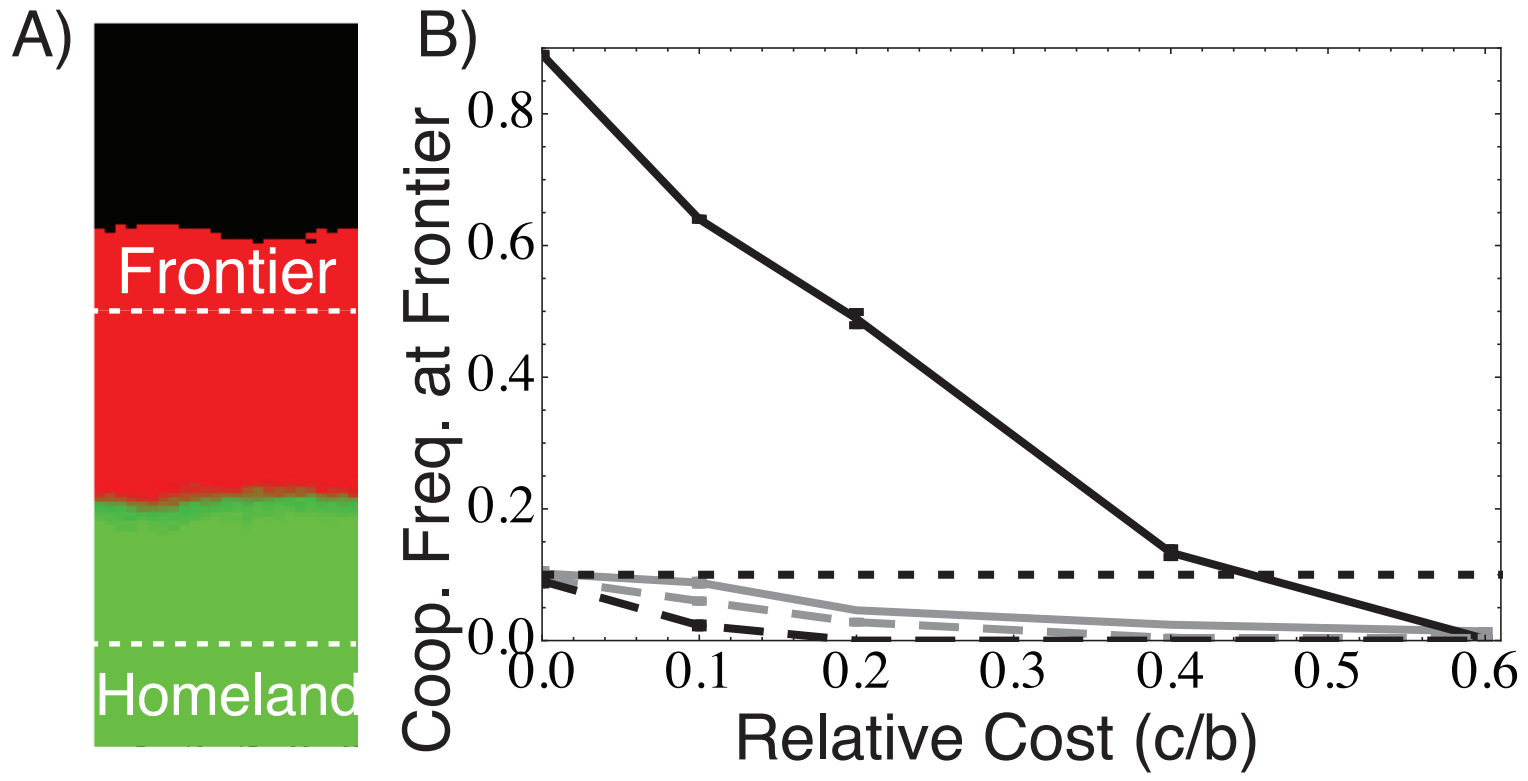

# Supplemental Information: "Spatial population expansion promotes the evolution of cooperation in an experimental Prisoner's Dilemma"

J. David Van Dyken*[12], Melanie Mueller[23], Keenan M.L. Mack[4], Michael M. Desai[123]
[1]Department of Organismic and Evolutionary Biology, Harvard University,
[2]FAS Center for Systems Biology, Harvard University
[3]Department of Physics, Harvard University
[4]University of Tennessee, Knoxville
*vandyken@fas.harvard.edu

## Supplemental Inventory

## Heuristic description of model and derivation of analytical results:

### Mechanism of selection for cooperation during range expansion

The fate of cooperators during a range expansion is determined by two phases, which we will take in turn. In Phase I, stochasticity at the frontier due to serial genetic bottlenecking causes local fixation of genotypes and loss of genetic diversity. Despite being selected against within subpopulations, cooperators can nonetheless fix at the frontier if stochastic effects overwhelm purifying selection. Once genotypes fix at the frontier, the expanding wave of individuals (the "density wave") will almost always travel faster than the "allele frequency wave" (Figure S1). To see this, note that in a population growing logistically with diffusion of individuals via dispersal into neighboring habitat sites, the speed of such a wave in steady-state has a known solution,

$$v_d = 2\sqrt{D\bar{W}} \tag{1a}$$

as does the speed of the traveling allele-frequency wave [1]

$$v_f = 2\sqrt{D\left(W_i - W_j\right)} \tag{1b}$$

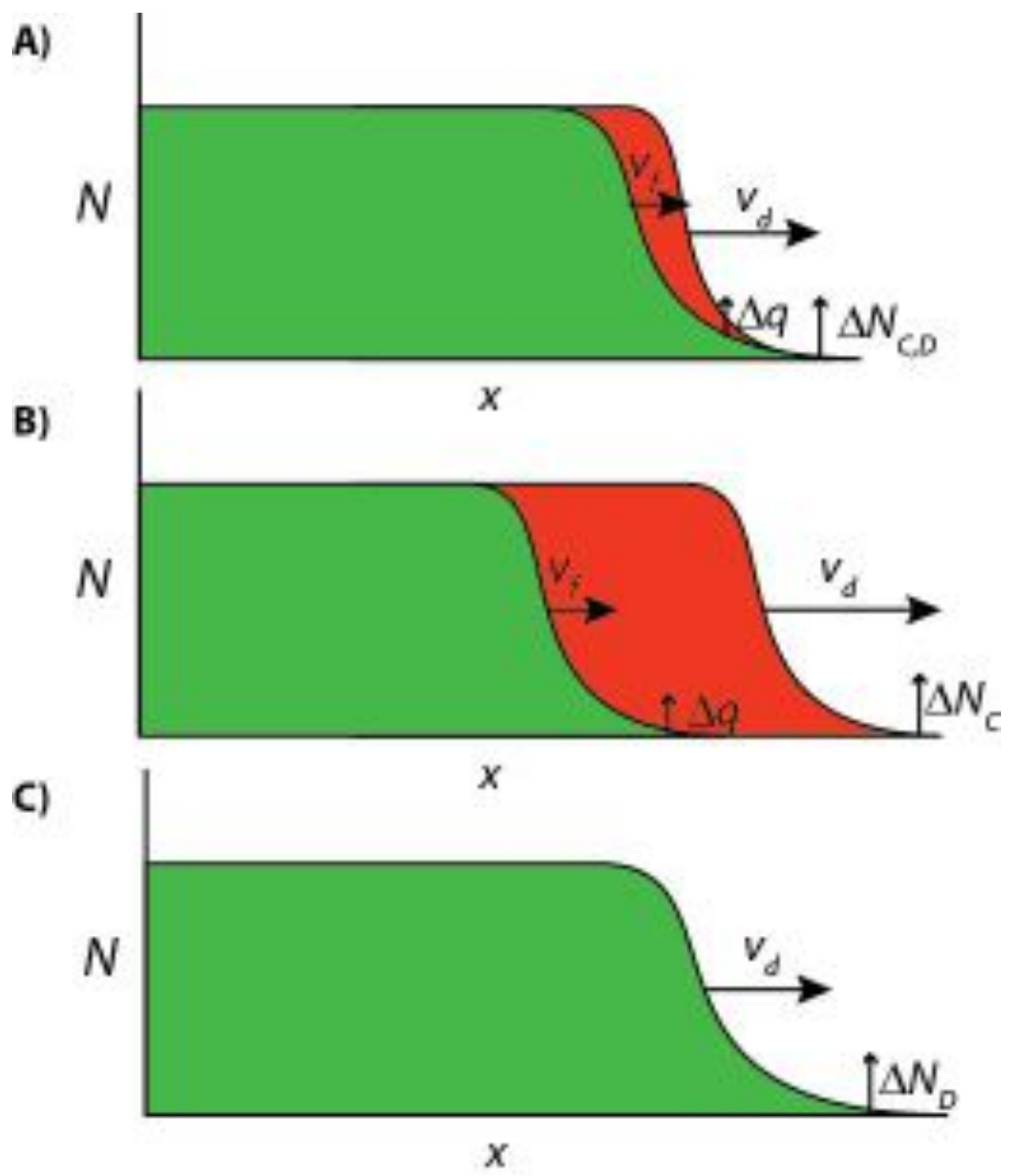

**Figure S1| Spatial expansion in one-dimension.** Population size, *N*, as a function of spatial coordinate, *x*, in a spatially expanding population. In the Prisoner's Dilemma, a mixed cooperator (red) and defector (green) front (**A**) will eventually resolve into a wave fixed for either cooperators (**B**) or defectors (**C**). Because cooperators lose in direct competition to defectors within each subpopulation (i.e., at each site *x*), outcome (B) requires that the stochastic effects of sampling at the leading edge overcome selection. However, once fixed, cooperators (or any other genotype) will outrun defectors, which advance in the trailing allele frequency wave. If defectors fix at the front (C), there is no trailing allele frequency wave because cooperators cannot invade defector subpopulations.

$D$ is the diffusion constant, $\bar{W}$ is the mean Malthusian fitness of a subpopulation (the maximal rate of increase), and $W_k$ is the fitness of genotype $k$. In the Prisoner's Dilemma game we can write the fitness of cooperators and defectors as,

$$W_C = W_0\left(1 + bp_C - c\right) \tag{2a}$$

$$W_D = W_0\left(1 + bp_C\right) \tag{2b}$$

where $W_0$ is the baseline reproductive rate, $b$ is the fitness benefit donated by cooperators, $c$ is the loss in fitness from cooperating and $p_C$ is the frequency of cooperators in a subpopulation (we assume that subpopulations are, by definition, well mixed in terms of social interactions, so that every individual has an equal chance of interaction with any other individual, and that choice of social partners is indiscriminate). From this, we have,

$$v_d = 2\sqrt{DW_0\left(1 + \left(b - c\right)p_C\right)} \tag{3a}$$

$$v_f = 2\sqrt{\left(\pm\right)DW_0c} \tag{3b}$$

where the $c$ is positive if cooperators are fixed at the front, and negative if defectors are fixed at the front. This means that defectors can deterministically invade subpopulations fixed for cooperators (Figure S1B), but cooperators cannot invade defectors (Figure S1C). So, we have 2 scenarios corresponding to Figures S1B,C:

*1) Cooperators stochastically fix at the frontier (Figure S1B):*

a) Density wave of cooperators proceeds at speed:

$$v_d = 2\sqrt{DW_0\left(1 + b - c\right)} \tag{4a}$$

b) And is trailed by a lagging frequency wave of cheaters at speed:

$$v_f = 2\sqrt{DW_0c} \tag{4b}$$

c) Cooperation is stable at the front, and will contintue to gain ground and increase in global frequency if:

$$\frac{1+b}{2} > c \tag{4c}$$

*2) Defectors stochastically and/or deterministically fix at the frontier (Figure S1B):*

a) Density wave of defectors proceeds at speed:

$$v_d = 2\sqrt{DW_0} \tag{5}$$

b) There cannot be a lagging frequency wave of cooperators, because its speed is not a real number.

c) Cooperation is eliminated.

Because of this, cooperators can only increase in frequency in a one-dimensional range expansion if they stochastically fix at the front. Thus, there is no positive force of selection promoting cooperation in the PD in one dimension.

Importantly, this analysis has thus far been restricted to a single spatial dimension. In reality, most range expansions will proceed along two spatial dimensions. Two-dimensional range expansions can undergo Phase II: competition among neighboring subpopulations at the frontier favoring genotypes with high productivity (Figure S2). Note that within subpopulations, selection favors genotypes with the highest *relative* fitness, which in the case of spite and selfishness (see below) actually causes a reduction in total reproductive output as these traits sweep to fixation. This is known as the "Tragedy of the Commons". However, at the frontier, expansion speed is determined by *absolute* fitness, generating a force of selection promoting genotypes that increase productivity. In the Prisoner's Dilemma, equations S4a,b tell us that subpopulations fixed for cooperators will travel faster than those fixed for defectors if,

$$b > c \tag{6}$$

Together, Phase I (genetic demixing) and Phase II ("survival of the fastest") interact to create a force promoting high productivity strategies such as cooperation (Figure S2). However, the specific parameter regimes in which this force can overcome selection within groups favoring defectors, which is necessary for Phase I, are not immediately clear. A more detailed theoretical analysis of the interactions between density and allele frequency waves in two dimensions will be required to provide specific conditions necessary for cooperation to evolve in expanding populations, taking into account both Phase I and II; this is an interesting topic for further work.

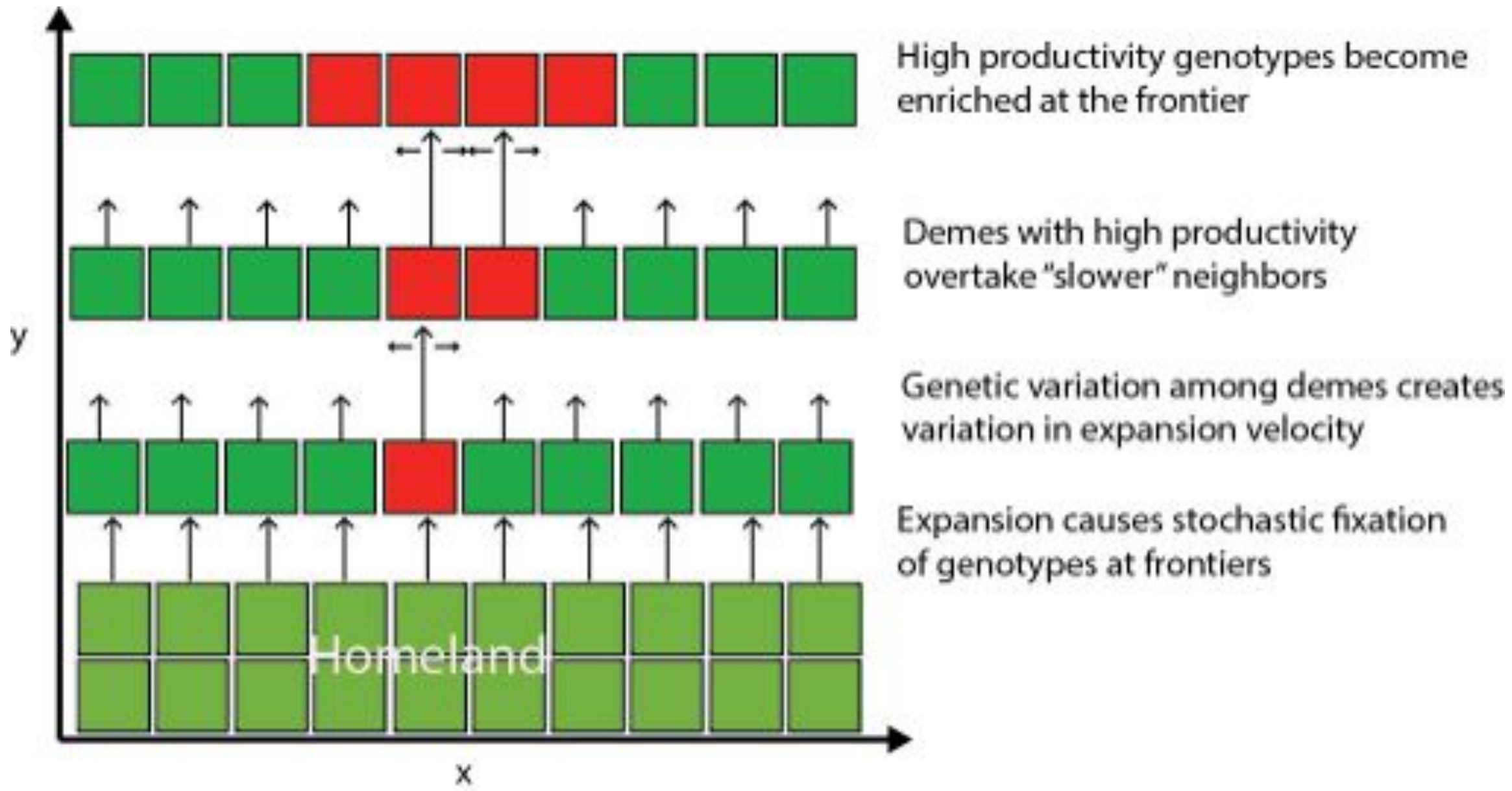

**Figure S2| Spatial expansion in two-dimensions allows selection for cooperation.** Populations expanding in two spatial dimensions, with each site at coordinate ($x$,$y$) representing a subpopulation connected to nearest neighbors by dispersal according to Kimura's stepping stone model [2]. A mixed homeland with rare cooperators will eventually demix upon expansion into subpopulations fixed for either cooperators (red) or defectors (green). Because subpopulations with cooperators expand faster than subpopulations of defectors, cooperators become enriched at the frontier by overtaking neighboring defector sectors.

**Extension to other social behaviors**

Spatial population expansions may influence the evolution of other social behaviors as well. Consider spite and selfishness, strategies that reduce mean population fitness, but may spread nonetheless. Spite occurs when an individual reduces its own personal fitness to harm others ($b < 0$, $c > 0$) [3, 4]. Despite counterselection within subpopulations (due to fitness cost, $c$), spite can nonetheless stochastically fix at the frontier of an expanding population. Once this happens, equation 4c tells us that spitefull genotypes will increase in frequency in one-dimensional populations as long as $c < (1 - b)/2$, which can be satisfied over a wide range of parameter space.

Selfish individuals increase their direct fitness while reducing the fitness of neighbors ($b$, $c < 0$). Selfish genotypes are more likely to fix at frontiers than non-selfish genotypes because of positive selection (direct benefit of magnitude, $c$), and will also establish an allele-frequency wave that will chase non-selfish genotypes that stochastically fix at the frontier.

Two-dimensional spatial expansions, however, select against both spite and selfishness. Lower productivity of spiteful and selfish subpopulations makes them vulnerable to being overtaken by neighboring subpopulations of non-spiteful or non-selfish strategies. A full exploration of this effect awaits further study.

**Experimental Procedures**

**Strains:** Strains were haploid (MATa) prototrophs with deletions of *mal11* and *mal12* genes, constructed from W303 background with *ADH*1 promoter-driven expression of the fluorescent markers ymCherry (cooperators) and ymCitrine (*suc2Δ* defectors).

**Experimental setup:** Overnight cultures grown in YPD were washed twice with sterile water, resuspended to a density of $2x10^8$ cells/mL as measured by Coulter Counter, and mixed in appropriate ratios quantified by FACS. *Stationary (non-expanding) competitions* were conducted in round-bottom 96-well culture plates with 128 uL of liquid media: YEP (1% yeast extract, 2% peptone) plus 2% filter sterilized sucrose and the appropriate concentration of cycloheximide, both added after autoclaving. Cycloheximide stocks were diluted in ethanol, filter sterilized and stored at -20 C until use. Wells were inoculated with 1uL of initial culture, and 1uL was passaged from each well into fresh media every 24 hours. Plates were incubated at 30 C on an orbital plate shaker at 1000 rpm. Strain frequencies were measured using FACS at days 0, 1, 2, and 6, with 3 replicates of each condition. *Range expansion competitions* were conducted on 7 mL agar media (same recipe as above plus 2% agar) in 6-well culture plates. The two center wells were left empty to avoid plate effects. 1 uL of initial culture was spotted onto the center of each well and plates were incubated at 30 C, with 3 replicates of each condition. For frequency analysis, 3 replicates from each condition were chosen at random and harvested at the appropriate time point by repeatedly pipetting 2 mL PBS until colony was completely detached from the agar and well mixed, then the culture was diluted appropriately for FACS analysis. *Growth rates* were conducted as with the range expansions, but conducted in individual petri plates on 12 mL of agar media.

**Image analysis:** Image analysis was performed with Matlab. Colony radii were determined from circle fits to the colony boundary, detected using edge detection or thresholding on the brightfield image of the colony. Sector boundaries were identified by edge detection in the fluorescent images. Each sector was assigned a color by comparing its average intensity to the average intensity of its neighboring sectors.

**Simulations:** Simulation were conducted on an *n*x*m* square lattice. Each site of the lattice contained a subpopulation of size $N_T$ undergoing logistic population growth with genotype-independent carrying capacity, *K*, non-overalapping generations, and growth rate of genotype *i* at local frequency $p_i$ following: $N_i' = (p_i + \Delta p_i)(N_T + \Delta N_T)$, where $\Delta N_T = W_0(1+(b-c)p_C)N_T(1-N_T/K)$, $\Delta p_i = p_i(1-p_i)(W_i - W_j)$, and $(W_C - W_D) = -c$. Stochasticity was introduced by first computing $N_i'$ as a real number, then using the non-integer part as the binomial probability of success over $N_i$ trials. The life cycle was as follows: cooperators produce a fitness benefit *b*, which is shared equally by all $N_i$ group-mates, at a personal fitness cost *c*, individuals then reproduce based on their fitness, die, and finally disperse. The number of migrants of each type was a binomial random draw with success probability equal to *m*. Simulations were initiated with a "homeland" population with each site at carrying capacity and a binomial random number of cooperators with mean frequency equal to $p_0$. In linear expansions, the homeland consisted of 10 rows of sites filling the bottom of a cylindrical lattice (boundaries were periodic only in the direction perpendicular to the direction of expansion). Radial expansions were initiated as a square block of demes in the center of a lattice. Scripts were written in Matlab with plots and analysis performed in Matlab and Mathematica.

**Additional Supplemental Figures**

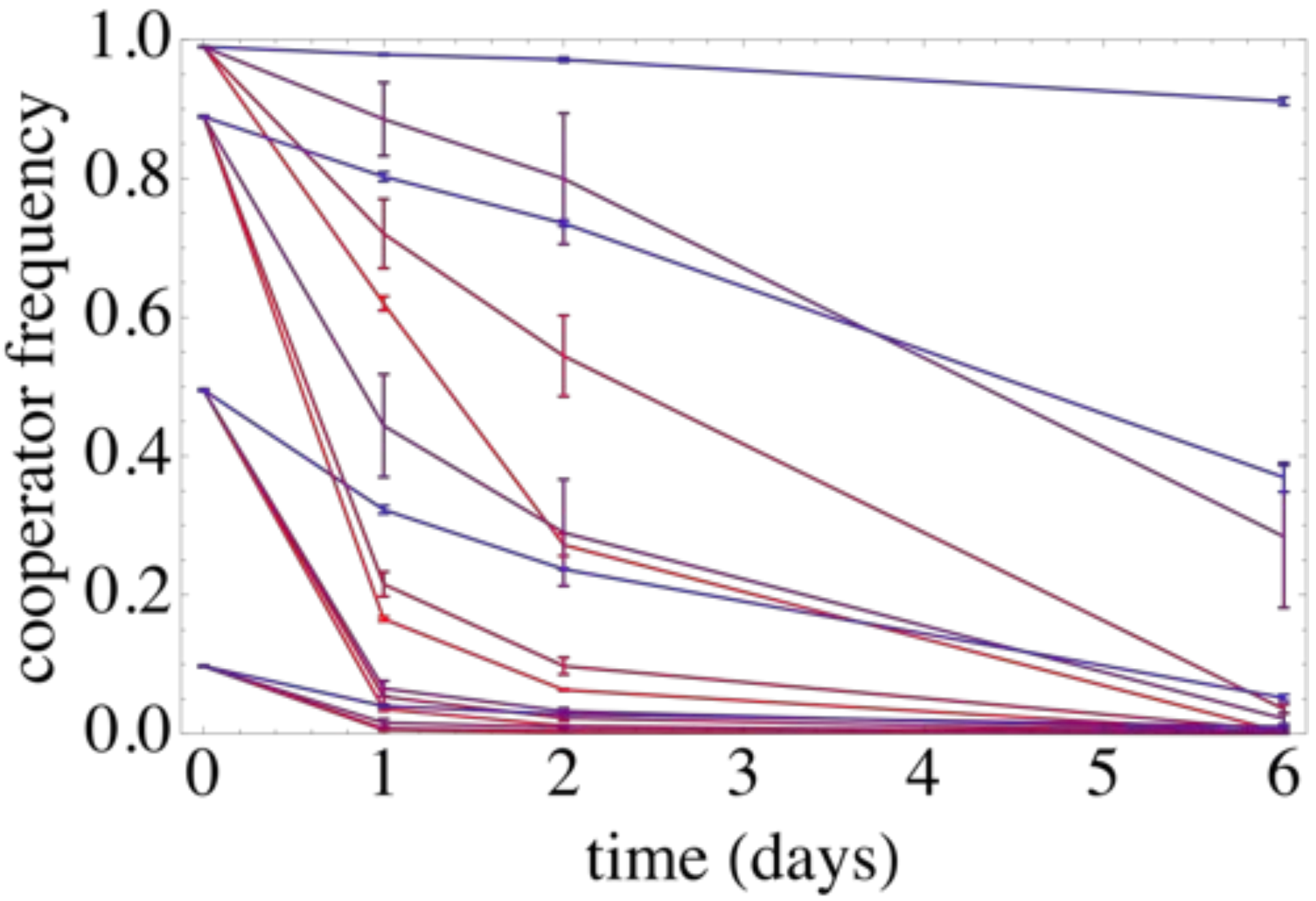

**Figure S3| Selection against cooperators in a stationary environment.** Competition between cooperator and defector strains in shaken liquid culture for a range of imposed costs (cycloheximide concentration varied from 50 (blue) to 150nM (red): 50 nM, 75 nM, 100 nM, 150nM). Strain frequencies measured by FACS. Any potential equilibrium between cooperators and defectors is below 0.01, making our system function as a Prisoner's Dilemma for all conditions considered: increase in cooperator frequency in our range expansions cannot be due to rare advantage.

**A)**

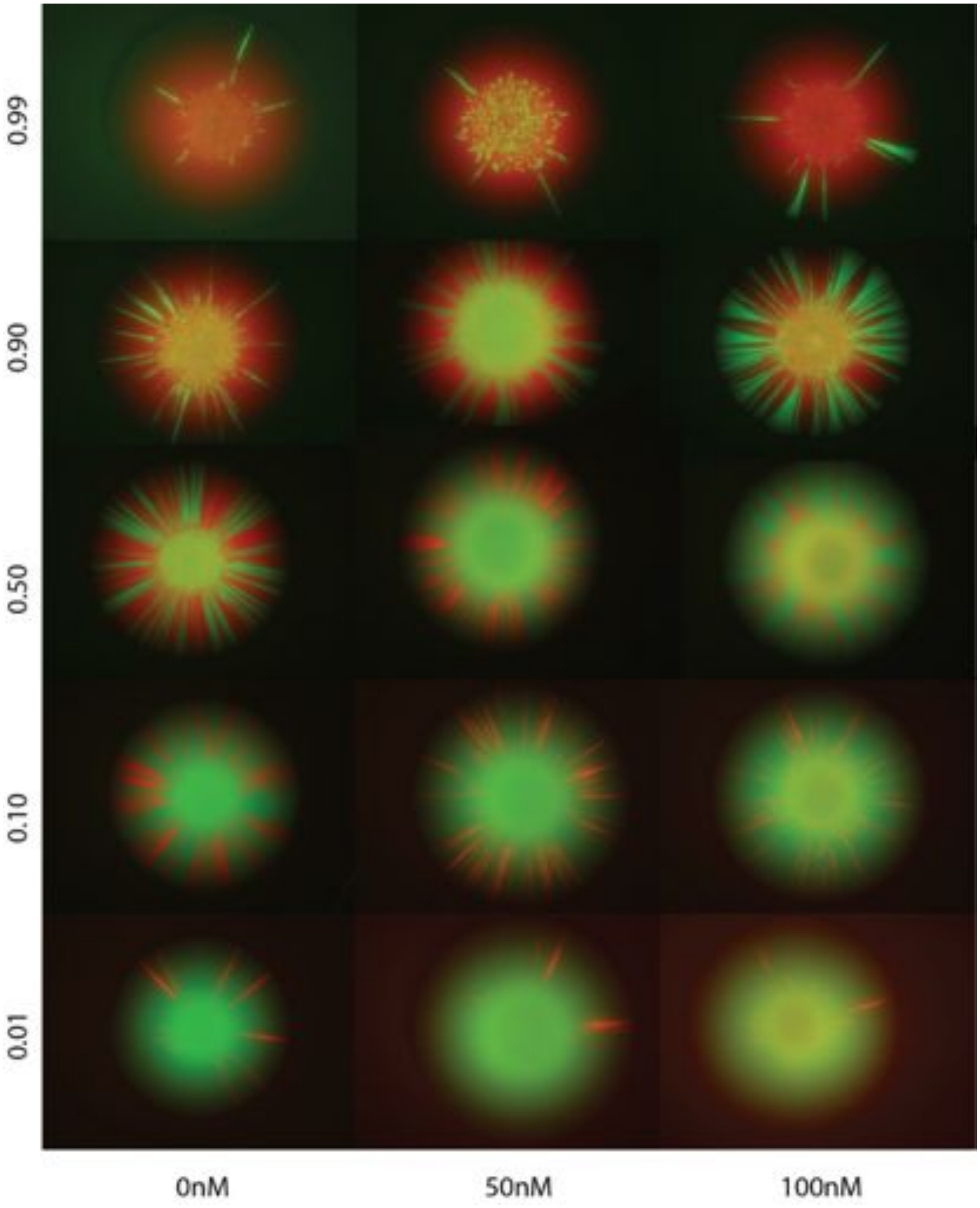

**B)**

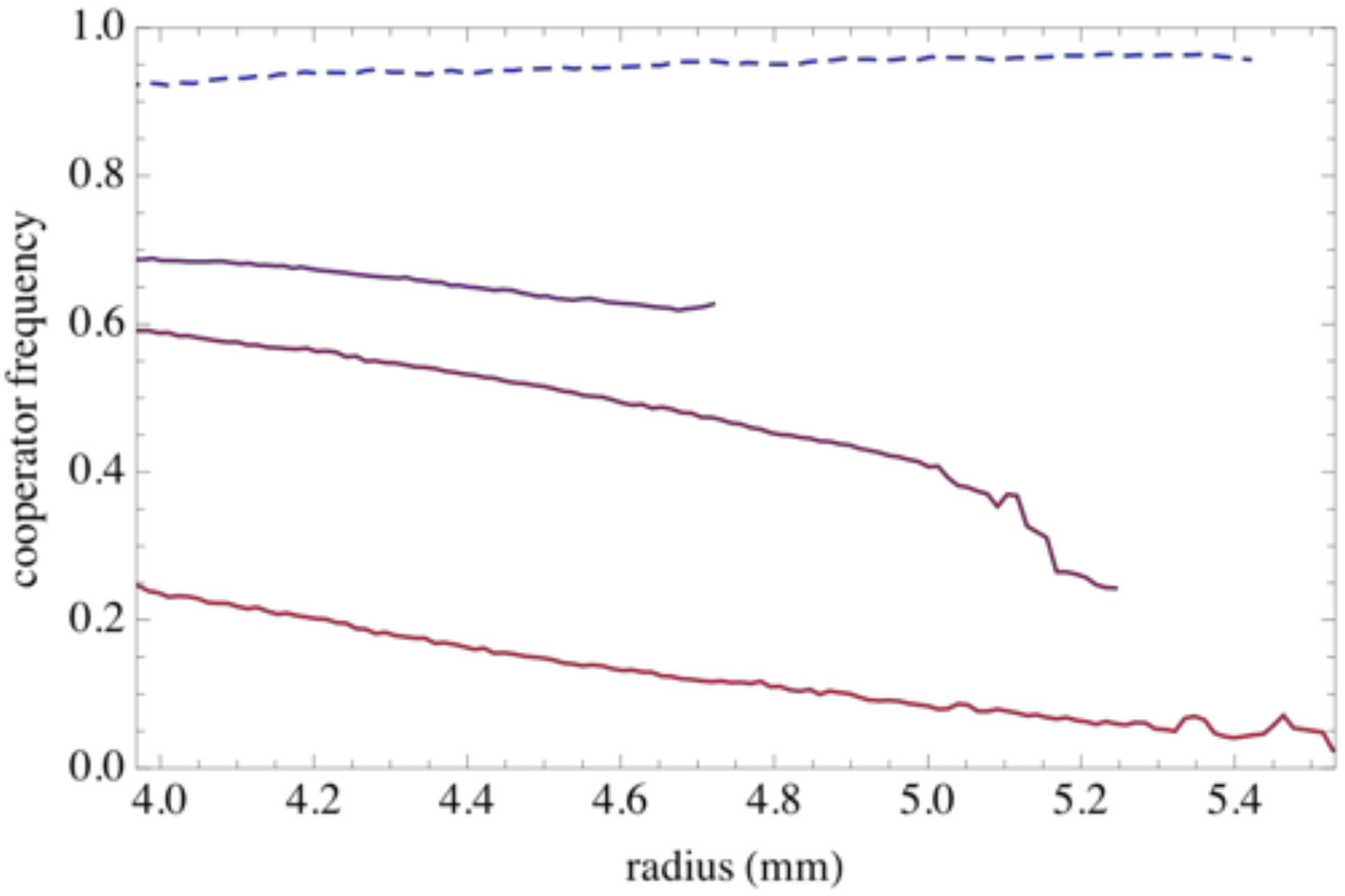

**Figure S4| Range expansions in glucose media controls. A)** Cooperator (red) and defector (green) strains competing in expanding colonies on glucose rich media (YEP + 2% glucose + 2% agar). The abundance of monosacharides and the absence of sucrose in the media render the coopertive phenotype, sucrose digestion, unnecessary. Thus, the benefit of cooperation is eliminated leaving only the cost of cooperation to distinguish strains. **B)** Image analysis of glucose controls for colonies with an initial frequency of cooperators of 0.90. Cycloheximide concentrations, from top to bottom: 0nM (dashed blue line), 75nM, 100nM and 150nM. Importantly, in these control experiments cooperators decline in frequency (from 90% in this case) when cycloheximide is applied, in contrast to the case when cooperation is beneficial in sucrose media (Figure 2E in main text). Note that the 0nM (dashed blue) line increases in frequency slightly from 90%, indicating a slight cost to cycloheximide resistance in the cheaters in the absence of cycloheximide.

## Supplemental References